\documentclass[pdflatex,sn-mathphys-num]{sn-jnl}
\usepackage{graphicx}
\usepackage{placeins}
\usepackage{multirow}
\usepackage{amsmath,amssymb,amsfonts}
\usepackage{amsthm}
\usepackage{mathrsfs}
\usepackage[title]{appendix}
\usepackage{xcolor}
\usepackage{textcomp}
\usepackage{booktabs}
\usepackage{mathtools}
\usepackage{bm}

\graphicspath{{./graphics/}}

\theoremstyle{thmstyleone}

\theoremstyle{thmstyletwo}

\theoremstyle{thmstylethree}

\begin{document}

\title[Residual detuning in laboratory-frame AJC dynamics]
{Residual detuning in the laboratory-frame
anti-Jaynes--Cummings model with a squeezed vacuum}

\author*[1]{\fnm{Onyango Stephen} \sur{Okeyo}}
\email{ookeyo@maseno.ac.ke; onyango@physik.hu-berlin.de}

\affil[1]{\orgdiv{Department of Physics and Materials Science},
\orgname{Maseno University},
\orgaddress{\street{Private Bag}, \postcode{40105},
\state{Maseno}, \country{Kenya}}}

\abstract{
The laboratory-frame anti-Jaynes--Cummings (AJC) interaction retains a
residual detuning $2f\lambda$ that is absent from the rotating-frame
model. We map two proposed remedies---a Kerr shift
$\chi(\hat a^\dagger\hat a)^2$ and collective Dicke coupling of $N$
two-level emitters---for a squeezed vacuum on that ladder ($r=1$,
$\langle n\rangle=\sinh^2 r\simeq 1.38$). All quoted contrasts are the
amplitude of the first turning point of the atomic ground-state
population, which coincides with the two-level formula
$\mathcal{C}=1/[1+(2f+\chi)^2]$ at $r=0$ to $10^{-9}$. Three results
follow.
(i)~A single Kerr strength never restores unit contrast at $r=1$;
the $r=0$ $n$-dependent shift $\chi(2n+1)$ cannot cancel $2f\lambda$
on every occupied Fock component.
(ii)~The exact $r=1$ contrast at $\chi=0$ is not reproduced by an
incoherent sum $\sum_n P_n(r)\,\mathcal{C}_n$ built from the $r=0$
two-level formula; pointwise deviations are several tenths.
(iii)~Collective coupling raises the contrast systematically. At
$f=5$ one finds $\mathcal{C}=0.140$ ($N=1$) and $\mathcal{C}=0.575$
($N=8$), above the unsqueezed value $8/[8+(2f)^2]=0.074$.
The $N=16$ point at this $f$ remains truncation-limited and is
not quoted to three digits.
The same $N\sim(2f)^2$ estimate for $\mathcal{C}=1/2$ places
trapped-ion values $f\sim 10^{2}$--$10^{3}$ outside the present
construction. The relevant platform is ultrastrong circuit QED with
$f\sim 1$--$10$. The calculation is a numerical control landscape,
not a new solvable limit.
}


\maketitle

\section{Introduction}

The anti-Jaynes--Cummings (AJC) interaction
$K^\dagger\sigma_++K\sigma_-$ is the counter-rotating counterpart
of the Jaynes--Cummings model~\cite{Jaynes1963,Shore1993, Rodriguez2005,Bocanegra2024} and couples
a two-level emitter to the creation part of a bosonic ladder. In a
rotating frame chosen to cancel the free evolution of field and atom,
that coupling is resonant. In the laboratory frame the same model
retains a residual detuning $2f\lambda$, where $\lambda$ is the vacuum
Rabi amplitude and $f=\omega/\lambda$~\cite{Mayero2023,Mayero2024}.
The detuning is a kinematical feature of the frame.

Throughout this work ``laboratory frame'' means the frame in which the
free field term $\omega\hat a^\dagger\hat a$ and the atomic term
$(\omega_0/2)\sigma_z$ are kept explicitly. In the rotating frame used
for trapped-ion sideband driving~\cite{Leibfried2003,Kienzler2015,Kienzler2017},
those terms are transformed away and $2f\lambda$ is absorbed into the
drive; it is not an observable there. The numbers below apply to the
laboratory-frame problem and therefore to ultrastrong circuit QED, where
counter-rotating terms remain and
$f\sim 1$--$10$~\cite{Blais2004,Blais2021,FornDiaz2019,Kockum2019}.

A Kerr term $\chi(\hat a^\dagger\hat a)^2$ shifts Fock level $n$ by
$\chi n^2$ and moves the effective detuning of the $n\leftrightarrow n+1$
AJC block to
\begin{equation}
  \Delta_n=2f\lambda+\chi(2n+1).
  \label{eq:Delta}
\end{equation}
The $r=0$ origin of Eq.~\eqref{eq:Delta} is given in
Appendix~\ref{app:kerr}.
A single $\chi$ cancels $\Delta_0$ when $\chi=-2f\lambda$, which is exact
for a vacuum. A squeezed vacuum occupies a broad even-$n$ packet, so the
same $\chi$ cannot cancel every $\Delta_n$. Independently, $N$ emitters
sharing one mode increase the bright-state Rabi frequency as
$\sqrt N$~\cite{Dicke1954,Tavis1968,Garraway2011}. For an unsqueezed
vacuum the resulting contrast is
\begin{equation}
  \mathcal{C}_{\mathrm{Dicke}}(N,f)=\frac{N}{N+(2f)^2}.
  \label{eq:dicke}
\end{equation}

Neither knob is conceptually new~\cite{Phoenix1991,Gora1992,Buzek1992}.
Laboratory-frame AJC dynamics with squeezed light were treated in
Refs.~\cite{Mayero2023,Mayero2024,Omolo2025QIP}; those papers did not
scan a Kerr compensator or a Dicke ladder. Related AJC studies map the
unsqueezed model onto JC by a duality or supersymmetry
transformation~\cite{Rodriguez2005,Bocanegra2024}; they do not
treat the squeezed operator $K$ in the laboratory frame. The question
here is narrower: in the laboratory frame, with a squeezed vacuum on
the ladder, how much contrast is recovered by Kerr compensation and by
collective coupling, and is the $r>0$ dynamics just an incoherent
average over Fock blocks? The answers are numerical. 
Section~\ref{sec:model} defines the laboratory-frame Hamiltonian,
the short-time contrast, and the numerical method.
Section~\ref{sec:results} reports the $r=0$ Kerr Lorentzian, the
failure of the incoherent average at $r=1$, and the Dicke scan.
Section~\ref{sec:scope} gives the atom-number estimate and the
circuit-QED window.
Appendix~\ref{app:kerr} derives the $r=0$ block $H_n$ and the
origin of $\Delta_n=2f\lambda+\chi(2n+1)$ in the parameters $d$
and $f$.
Appendix~\ref{sec:appendix} records the breaking of
$\overline N$ at $r\neq 0$.
Appendix~\ref{app:num} is the truncation table.

\section{Model}
\label{sec:model}

\subsection{Laboratory-frame Hamiltonian}

One emitter and one bosonic mode are described by
\begin{equation}
  H=f\lambda\,\hat n+\frac{(d+f)\lambda}{2}\sigma_z
  +\frac{\Omega}{2}\bigl(K^\dagger\sigma_++K\sigma_-\bigr)
  +\chi\,\hat n^2,
  \label{eq:H}
\end{equation}
with $\Omega=\lambda$, $d=0$, $\hat n=\hat a^\dagger\hat a$, and
$K=SaS^\dagger$, where $S$ is the squeeze
operator~\cite{Walls1983,Loudon1987}. We take $r=1$, $\phi=0$, for
which $\langle n\rangle=\sinh^2 r\simeq 1.381$ and only even photon
numbers are occupied. The AJC (plus) ladder of Eq.~\eqref{eq:H} is
compared with the JC (minus) replacement $K\leftrightarrow K^\dagger$
where stated. The alternative Kerr $\chi\,\hat a^{\dagger 2}\hat a^2
=\chi\hat n(\hat n-1)$ does not cancel $\Delta_0$ at $\chi=-2f\lambda$;
it is a consistency check only and is not used for production numbers.

Equation~\eqref{eq:H} differs from the laboratory-frame AJC model of
Refs.~\cite{Mayero2023,Mayero2024}, where the interaction is
$a\sigma_-+a^\dagger\sigma_+$. In that case
$\overline N=a^\dagger a+\sigma_-\sigma_+$ is conserved at every $r$,
because the interaction uses unsqueezed operators. Here $K$ mixes $a$
and $a^\dagger$, and the free term is $a^\dagger a$, not $K^\dagger K$,
so the conservation is broken for $r\neq 0$
(Appendix~\ref{sec:appendix}). Full truncated diagonalization is then
required; that is what is done below.

For $N$ emitters we replace $\sigma_z/2\to J_z$ and
$\sigma_\pm\to J_\pm$ in the Dicke subspace of total spin $J=N/2$,
\begin{equation}
  H_N=f\lambda\,\hat n+(d+f)\lambda\,J_z
  +\frac{\Omega}{2}\bigl(K^\dagger J_++KJ_-\bigr)+\chi\hat n^2.
  \label{eq:HN}
\end{equation}
The initial atomic state is $|J,-J\rangle$. Equation~\eqref{eq:HN} at
$N=1$ reproduces Eq.~\eqref{eq:H}.

\subsection{Contrast}

Let $P_\downarrow(t)$ be the population of the atomic ground state
(the Dicke ground state when $N>1$). The published contrast is
\begin{equation}
  \mathcal{C}=\bigl|P_\downarrow(0)-P_\downarrow(t_\ast)\bigr|,
  \label{eq:C}
\end{equation}
where $t_\ast$ is the first extremum of $P_\downarrow(t)$, refined by a
local cubic spline. Global $\max P-\min P$ over a long window is not
used: on a collapse--revival trace~\cite{Eberly1980} it mixes later
beats into the figure of merit and is less stable under Fock
truncation. A single-frequency cosine fit on a fixed window
$\lambda t\le 8$ was tested and discarded. It underestimates the
solvable $r=0$ limits and inverts the $N$-scaling of the Dicke scan.

At $r=0$ the AJC contrast is
\begin{equation}
  \mathcal{C}_{r=0}(f,\chi)=\frac{1}{1+(2f+\chi)^2}
  \label{eq:lorentz}
\end{equation}
in units $\lambda=1$. If $K$ assigned only an $n$-dependent Rabi
frequency to a diagonal mixture, the $r>0$ contrast would be
\begin{equation}
  \bar{\mathcal{C}}=\sum_{n}P_n(r)\,
  \frac{4V_n^2}{4V_n^2+[2f\lambda+\chi(2n+1)]^2},
  \label{eq:inc}
\end{equation}
with $V_n=\Omega/2$ or $V_n=(\Omega/2)\sqrt{n+1}$.
Section~\ref{sec:inc} tests Eq.~\eqref{eq:inc} against the exact
evolution.

\subsection{Numerical method}

The squeeze operator is the matrix exponential of
$\tfrac12(\zeta a^{\dagger 2}-\zeta^* a^2)$ in a Fock basis truncated at
$n_{\max}$. Single-atom production runs use $n_{\max}=200$
($\dim=402$). Settled Dicke cells in Table~\ref{tab:chi0} use
$n_{\max}=160$; the two-digit $f\ge 1$ entries are bounds from the
same cutoff walk (Appendix~\ref{app:num}).
Time evolution is spectral on $\lambda t\in[0,40]$ with
$8\times 10^3$ points.

The squeezed-ladder matrix elements satisfy
$|\langle\zeta,n+1|K^\dagger|\zeta,n\rangle|=\sqrt{n+1}$ to a maximum
error $7.6\times 10^{-15}$ through $n=11$. At $r=0$ the numeric
contrast reproduces Eq.~\eqref{eq:lorentz} to $1.5\times 10^{-9}$ on a
13-point $\chi$ grid at $f=0.5$, and gives $\mathcal{C}=0.2000$
($f=1$, $\chi=0$) and $\mathcal{C}=1.0000$ ($f=1$, $\chi=-2$).
The $N=1$ Dicke reduction agrees with the single-atom code to four
digits at $r=1$, $f=1$, $\chi=0$ ($\mathcal{C}=0.3792$). Short-time
contrast at $(f,\chi)=(0.2,-0.4)$, $(2,0)$ and $(1,-0.3)$ changes by
less than $1.5\times 10^{-4}$ between $n_{\max}=160$ and $200$
(Appendix~\ref{app:num}).

\section{Results}
\label{sec:results}

\subsection{Vacuum AJC and the Kerr Lorentzian}

\begin{figure}[h]
\centering
\includegraphics[width=\columnwidth]{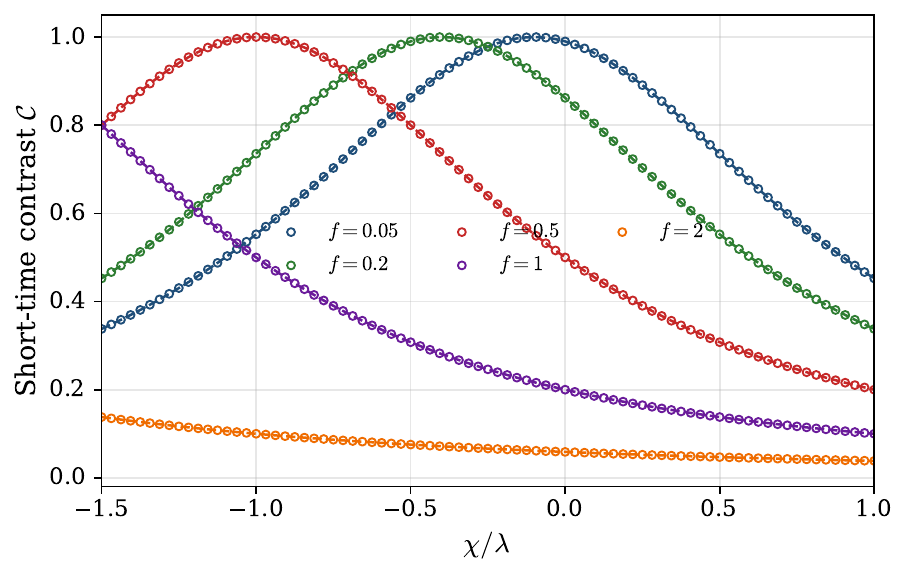}
\caption{Short-time AJC contrast versus Kerr strength at $r=0$.
Open circles: numerical diagonalization, $n_{\max}=200$.
Dashed: Eq.~\eqref{eq:lorentz}. Five values of $f$ are shown.
The Lorentzian peak at $\chi=-2f$ lies inside the plotted window
only for $f\le 0.5$; for $f=1$ and $f=2$ it sits at
$\chi=-2$ and $\chi=-4$. Units $\lambda=1$.}
\label{fig:r0}
\end{figure}

Figure~\ref{fig:r0} is the $r=0$ test of Eq.~\eqref{eq:H}. Open
circles are the numerical short-time contrast; dashed curves are
Eq.~\eqref{eq:lorentz}. The peak sits at $\chi=-2f$ and reaches
unity, as required.

\subsection{Squeezed vacuum at vanishing Kerr}
\label{sec:chi0}

\begin{figure}[htbp]
\centering
\includegraphics[width=\columnwidth]{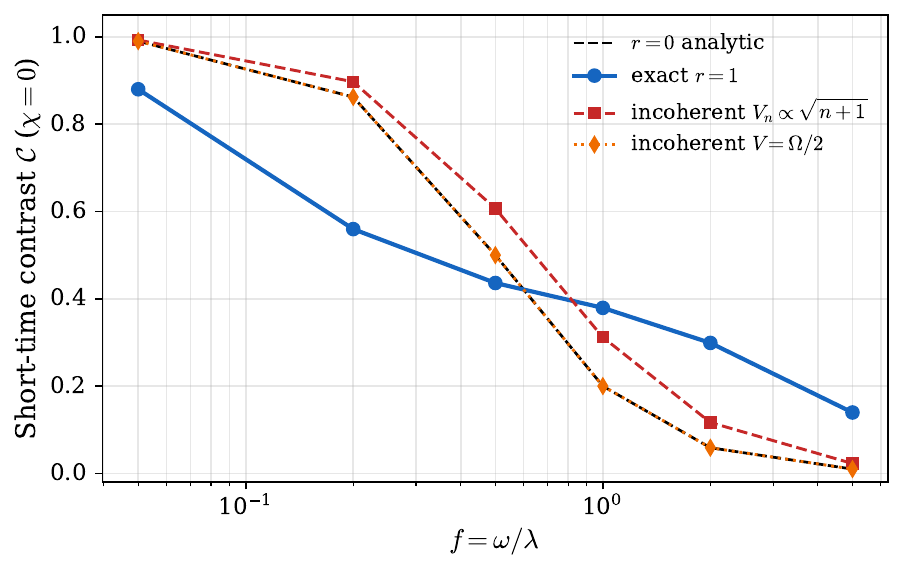}
\caption{Short-time AJC contrast at $\chi=0$ versus $f$.
Filled circles: exact $r=1$ evolution.
Squares (dashed): incoherent average with
$V_n=(\Omega/2)\sqrt{n+1}$.
Diamonds (dotted): incoherent average with $V=\Omega/2$.
Black dashed: $r=0$ analytic Lorentzian,
Eq.~\eqref{eq:lorentz}.}
\label{fig:inc}
\end{figure}

Table~\ref{tab:chi0} and Fig.~\ref{fig:inc} collect the $\chi=0$
landscape. For $r=0$ the numbers are Eq.~\eqref{eq:lorentz}. For $r=1$
the exact contrast lies below the vacuum value at small $f$ and above
it at large $f$. The crossing is between $f=0.5$ and $f=1$. JC
evolution at $r=0$ remains at unit contrast for every $f$ in the table.

\begin{table}[t]
\centering
\caption{Short-time contrast at $\chi=d=0$, $n_{\max}=200$
(single atom). Dicke values use $n_{\max}=160$ where $C$ and
$t_\ast$ settle ($f\le 0.5$, and $N=8$ at $f=5$).
Entries at $f\ge 1$ except $N=8$, $f=5$ are cutoff bounds
from Appendix~\ref{app:num}, not production values.
The inequality $\mathcal{C}_{N=16}(f=5)\gg 0.138$ holds at every
$n_{\max}$ in that appendix. AJC $r=0$ coincides with
$1/[1+(2f)^2]$ to the printed precision.}
\label{tab:chi0}
\begin{tabular}{@{}lcccccc@{}}
\toprule
$f$ & AJC $r=0$ & AJC $r=1$ & inc.\ $V_n$ & inc.\ $V$ & $N=8$ & $N=16$ \\
\midrule
$0.05$ & $0.990$ & $0.880$ & $0.993$ & $0.990$ & $0.983$ & $0.993$ \\
$0.20$ & $0.862$ & $0.560$ & $0.897$ & $0.862$ & $0.947$ & $0.978$ \\
$0.50$ & $0.500$ & $0.436$ & $0.607$ & $0.500$ & $0.892$ & $0.953$ \\
$1.00$ & $0.200$ & $0.379$ & $0.312$ & $0.200$ & $0.90$ & $0.93$ \\
$2.00$ & $0.059$ & $0.299$ & $0.117$ & $0.059$ & $0.86$ & $0.96$ \\
$5.00$ & $0.010$ & $0.140$ & $0.023$ & $0.010$ & $0.575$ & $0.78$ \\
\bottomrule
\end{tabular}
\end{table}

\begin{figure}[h]
\centering
\includegraphics[width=\columnwidth]{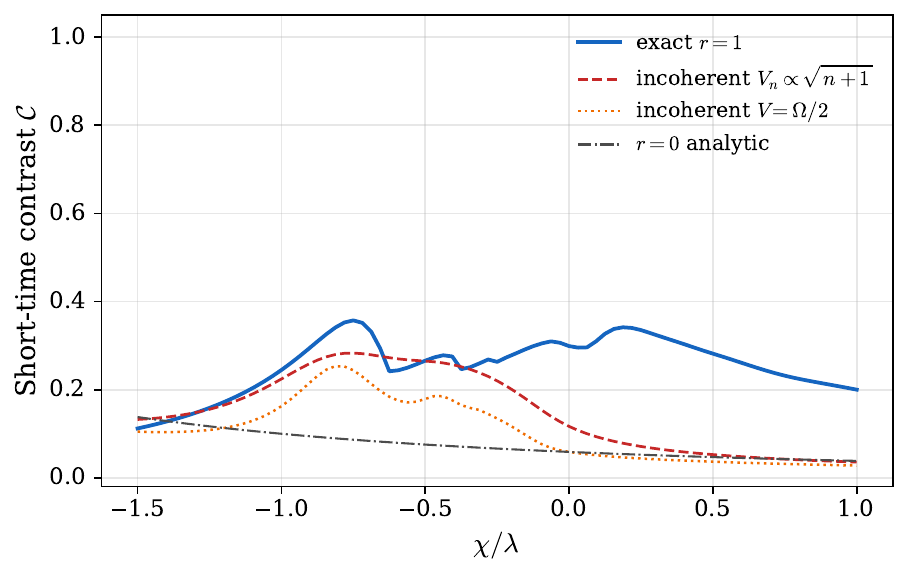}
\caption{Short-time contrast versus $\chi$ at $f=2$, $r=1$.
Solid: exact evolution. Dashed and dotted: the two
incoherent averages of Eq.~\eqref{eq:inc}.
Dash-dotted: $r=0$ analytic curve.
Diagnostic slice only; no $\chi$ optimum is extracted.}
\label{fig:kerrf2}
\end{figure}

\subsection{Failure of the incoherent average}
\label{sec:inc}

Figure~\ref{fig:inc} compares the exact $r=1$ contrast at $\chi=0$ with
both versions of Eq.~\eqref{eq:inc}. The two averages track the vacuum
Lorentzian more closely than they track the exact curve. At $f=1$ the
exact value is $0.379$ against $0.312$ (Fock coupling) and $0.200$
(vacuum coupling). At $f=2$ the corresponding numbers are $0.299$,
$0.117$ and $0.059$. The discrepancy shows that a diagonal mixture of
$r=0$ blocks is not the $r=1$ dynamics. The laboratory-frame free term
does not commute with $K^\dagger K$, so Fock coherences in the squeezed
vacuum and the mixing of photon-number sectors both survive. The present
test does not separate those two effects.

A full $\chi$ scan at $r=1$ is multi-chromatic. The first turning point
then jumps between incommensurate beat notes as $\chi$ is varied, and
$\mathcal{C}(\chi)$ is not a smooth Lorentzian. We therefore do not
quote a Kerr optimum at $r=1$.
Figure~\ref{fig:kerrf2} is a representative slice at $f=2$.
Within $|\chi/\lambda|\le 0.3$, the gain
relative to $\chi=0$ is modest and not robust under a change of contrast
definition. Kerr compensation is exact at $r=0$ and incomplete at
$r=1$.

\begin{figure}[htbp]
\centering
\includegraphics[width=\columnwidth]{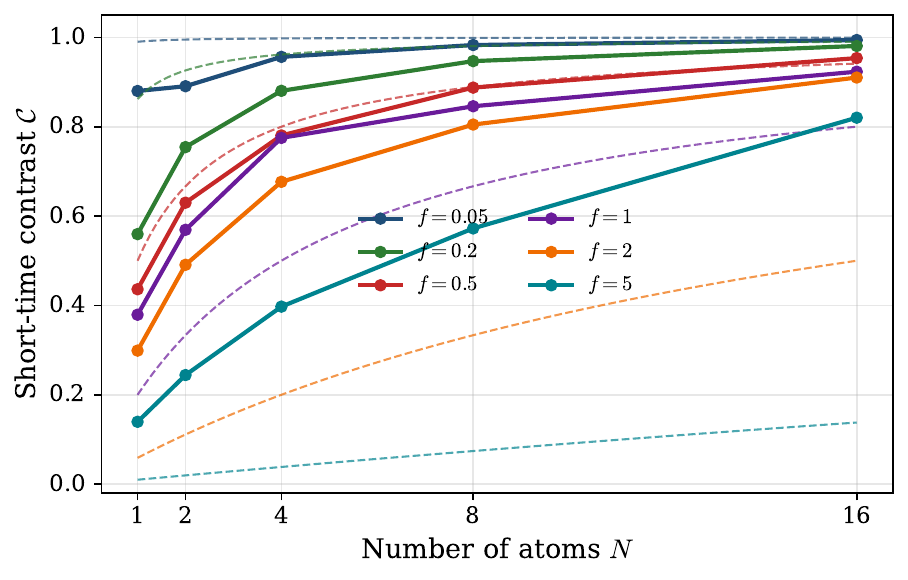}
\caption{Short-time AJC contrast versus atom number at $r=1$,
$\chi=0$. Solid: exact Dicke evolution for the six values of
$f$ in Table~\ref{tab:chi0}. Dashed: unsqueezed formula
Eq.~\eqref{eq:dicke}. The $N=8$ and $N=16$ points at $f\ge 1$,
except $N=8$ at $f=5$, are truncation-limited
(Appendix~\ref{app:num}).}
\label{fig:multi}
\end{figure}

\begin{figure}[htbp]
\centering
\includegraphics[width=\columnwidth]{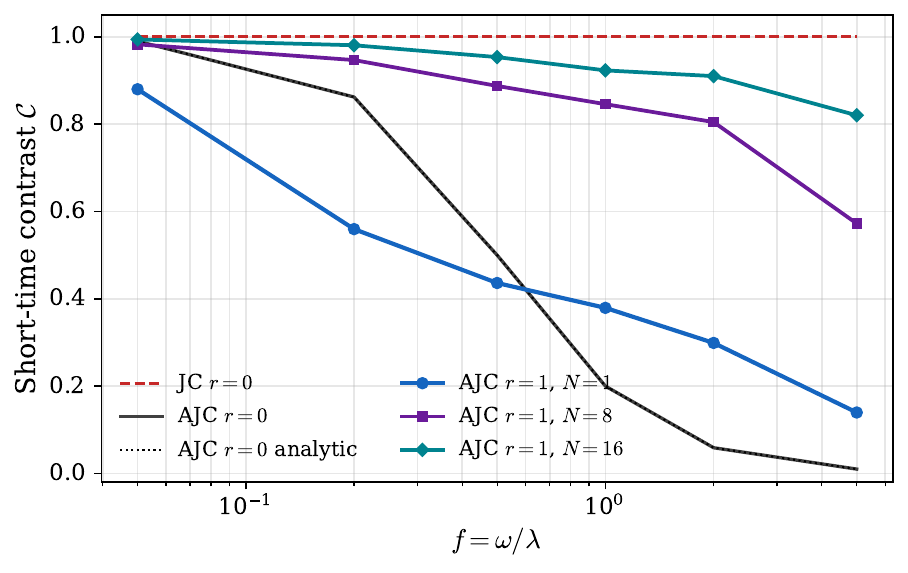}
\caption{Short-time contrast versus $f$ at $d=\chi=0$.
Red dashed: JC at $r=0$ (unit contrast).
Black: AJC at $r=0$, numeric and analytic.
Blue, purple, teal: AJC at $r=1$ for $N=1$, $8$, and $16$
(Table~\ref{tab:chi0}).
The $N=8$ and $N=16$ points at $f\ge 1$, except $N=8$ at $f=5$,
are truncation-limited (Appendix~\ref{app:num}).}
\label{fig:land}
\end{figure}

\begin{figure}[h]
\centering
\includegraphics[width=0.9\columnwidth]{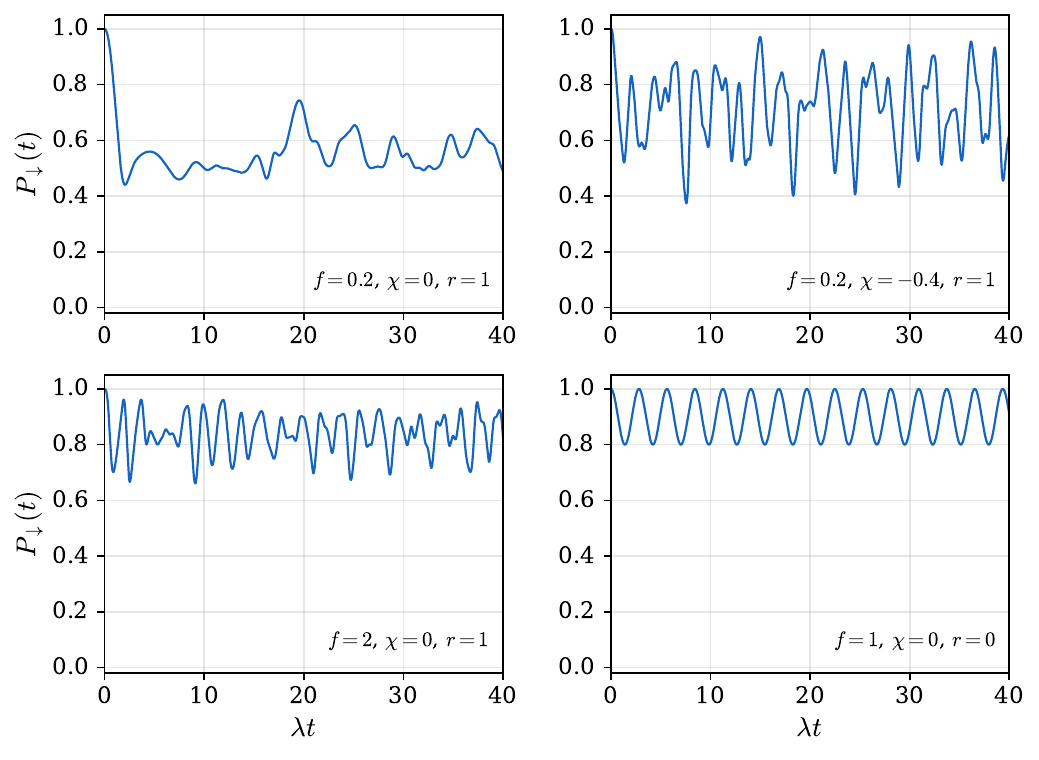}
\caption{Atomic ground-state population versus $\lambda t$.
Top left: $f=0.2$, $\chi=0$, $r=1$.
Top right: $f=0.2$, $\chi=-0.4$, $r=1$.
Bottom left: $f=2$, $\chi=0$, $r=1$.
Bottom right: $f=1$, $\chi=0$, $r=0$, amplitude $0.200$.
Contrast is $|P(0)-P(t_\ast)|$ at the first turning point;
later collapse--revival structure is not used.}
\label{fig:tr}
\end{figure}

\begin{figure}[h]
\centering
\includegraphics[width=0.9\columnwidth]{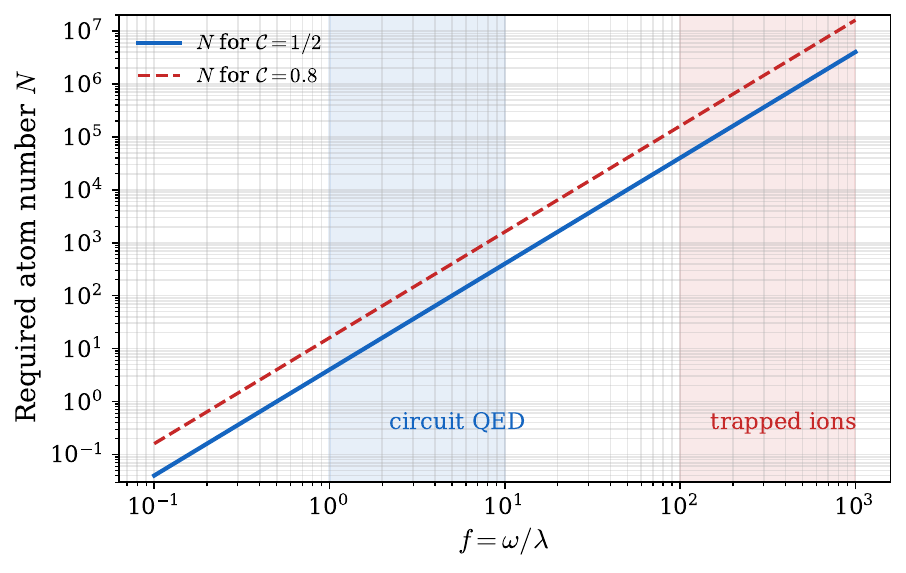}
\caption{Atom number implied by the unsqueezed estimate
Eq.~\eqref{eq:Nreq} for target contrasts $\mathcal{C}=1/2$
(solid) and $\mathcal{C}=0.8$ (dashed).
Blue band: circuit-QED window $f\sim 1$--$10$ used in this work.
Red band: trapped-ion window $f\sim 10^{2}$--$10^{3}$, where
$2f\lambda$ is removed by the rotating frame.}
\label{fig:N}
\end{figure}

\subsection{Collective enhancement}
\label{sec:dicke}

Figure~\ref{fig:multi} and Fig.~\ref{fig:land}, together with the last
two columns of Table~\ref{tab:chi0}, are the Dicke scan at $\chi=0$,
$r=1$. Dashed curves are Eq.~\eqref{eq:dicke}. At small residual
detuning ($f\le 0.2$) the exact contrast lies below that formula. At
large detuning ($f\ge 1$) it lies above, and the gap widens with $f$.
At $f=5$, $N=8$ one has $\mathcal{C}=0.575$ against
$8/[8+(2f)^2]=0.074$.
At $N=16$ the contrast stays well above $16/[16+(2f)^2]=0.138$
but $C$ and $t_\ast$ still move at $n_{\max}=160$
($\mathcal{C}\simeq 0.78$); that cell is not quoted to three digits.
Additional AJC channels opened by the squeezed ladder more than
compensate the residual detuning once the collective Rabi frequency
is large.
Equation~\eqref{eq:dicke} is a reference curve, not a fit to the
squeezed data.
At $f\ge 1$, except $N=8$ at $f=5$, the first turning point
changes identity when $n_{\max}$ is raised
(Appendix~\ref{app:num}).

\subsection{Time traces}

Figure~\ref{fig:tr} is the definition of Eq.~\eqref{eq:C}. The
bottom-right panel is vacuum AJC at $f=1$: a clean oscillation of
amplitude $0.200$, matching $1/(1+4)$. The other three panels are
squeezed packets. Later collapse--revival structure~\cite{Eberly1980}
is not used.

\section{Experimental scope}
\label{sec:scope}
\FloatBarrier

Equation~\eqref{eq:dicke} rearranged for a target contrast $C$ reads
\begin{equation}
  N=\frac{C(2f)^2}{1-C}.
  \label{eq:Nreq}
\end{equation}
Figure~\ref{fig:N} plots this for $C=1/2$ and $C=0.8$. At $C=1/2$ one
needs $N=(2f)^2$: $N=100$ at $f=5$, $N=400$ at $f=10$,
$N=4\times 10^4$ at $f=100$, and $N=4\times 10^6$ at $f=10^3$.
Trapped-ion sideband experiments operate at $f\sim 10^{2}$--$10^{3}$
and already implement the squeezed ladder in a rotating
frame~\cite{Leibfried2003,Kienzler2015,Kienzler2017}. Compensating the
laboratory-frame residual at those values of $f$ with collective
coupling on one mode is not realistic. The present numbers apply to
circuit-QED devices in the ultrastrong-coupling window
$f\sim 1$--$10$~\cite{Blais2004,Blais2021,FornDiaz2019,
Kockum2019,Dodonov2019}.
Even there a Kerr strength $|\chi|\sim\lambda$ strains the two-level
truncation of a transmon~\cite{Koch2007} and mixes $|2\rangle$ into
Eq.~\eqref{eq:HN}. The published spin-$1/2$ Dicke model does not
include that leakage.
The metric itself is a closed-system first turning point. Weak cavity
loss or qubit dephasing with $\kappa,\gamma\ll\lambda$ would reduce
$\mathcal{C}$ but is not expected to move $t_\ast$ out of the first
half-cycle used here. That statement is untested.

\section{Conclusions}
\label{sec:concl}

A squeezed vacuum on the laboratory-frame AJC ladder cannot be
detuning-compensated by one Kerr strength. The $r=0$ cancellation
$\chi=-2f\lambda$ is exact for the vacuum block and fails once the
packet occupies several even Fock components. The same dynamics is
not an incoherent sum over those blocks: at $\chi=0$ the exact
$r=1$ contrast differs from both versions of
Eq.~\eqref{eq:inc} by several tenths. Full truncated
diagonalization is required.

Collective coupling is the knob that still works at large residual
detuning. At $f=5$, eight emitters raise the short-time contrast
from $0.140$ to $0.575$, above the unsqueezed bright-state value
$0.074$. The $N=16$ cell at this $f$ remains truncation-limited.
The estimate $N\sim(2f)^2$ for $\mathcal{C}=1/2$ puts trapped-ion
values $f\sim 10^{2}$--$10^{3}$ outside the construction. The
relevant platform is ultrastrong circuit QED with $f\sim 1$--$10$,
where counter-rotating terms are kept in the laboratory frame.
Even there a Kerr strength $|\chi|\sim\lambda$ strains the
two-level truncation of a transmon, so the published spin-$1/2$
Dicke model is only a first estimate.

Two extensions follow directly. First, the $N=16$ walk at $f\ge 1$
must be pushed past $n_{\max}=160$ until $t_\ast$ stops jumping;
only then can that column be quoted to three digits. Second, the
closed-system first turning point should be retested with weak
cavity loss and qubit dephasing $\kappa,\gamma\ll\lambda$, and
with a transmon ladder that includes $|2\rangle$. Those runs
decide whether the contrast gain reported here survives in a
device. The present work is a numerical map of a known Hamiltonian,
not a new integrable limit.

\backmatter

\bmhead{Acknowledgments}
The author thanks the Department of Physics and Materials Science,
Maseno University, for support.


\section*{Declarations}
\noindent
Not applicable.

\begin{appendices}

\section{Kerr shift, $d$, and $f$ at $r=0$}
\label{app:kerr}

This appendix records the $2\times 2$ AJC block at vanishing squeezing,
where $K=a$. It is the only case in which Eq.~\eqref{eq:Delta} is an
exact eigenvalue difference. At $r\neq 0$ the block structure is lost
(Appendix~\ref{sec:appendix}) and the numbers in the main text are
obtained by full diagonalization.

Units $\hbar=1$. The dimensionless parameters in Eq.~\eqref{eq:H} are
\begin{equation}
  d\equiv\frac{\omega_0-\omega}{\lambda},\qquad
  f\equiv\frac{\omega}{\lambda},
\end{equation}
so $\omega=f\lambda$ and $\omega_0=(d+f)\lambda$. The JC difference
frequency is $\lambda d$. The AJC sum frequency is
$\omega_0+\omega=\lambda(d+2f)$. At resonance $d=0$ the latter
reduces to the residual $2f\lambda$.

On the invariant subspace
$\{|n,\downarrow\rangle,|n+1,\uparrow\rangle\}$ the laboratory-frame
AJC Hamiltonian with a diagonal Kerr term is

\begin{equation}
  H_n=
  \begin{pmatrix}
    f\lambda\, n+\chi K_n-\dfrac{(d+f)\lambda}{2}
    & \dfrac{\lambda}{2}\sqrt{n+1}\\[6pt]
    \dfrac{\lambda}{2}\sqrt{n+1}
    & f\lambda\,(n+1)+\chi K_{n+1}+\dfrac{(d+f)\lambda}{2}
  \end{pmatrix},
\end{equation}

where $K_m=m^2$ for $(\hat a^\dagger\hat a)^2$ and
$K_m=m(m-1)$ for $\hat a^{\dagger 2}\hat a^2$. The difference of
the diagonal entries is the effective detuning
\begin{equation}
  \Delta_n=
  \begin{cases}
    \lambda(d+2f)+\chi(2n+1)
    & \text{for }(\hat a^\dagger\hat a)^2,\\[2pt]
    \lambda(d+2f)+2\chi n
    & \text{for }\hat a^{\dagger 2}\hat a^2.
  \end{cases}
\end{equation}
With $\Omega=\lambda$ the off-diagonal element is
$V_n=\lambda\sqrt{n+1}/2$, so the Rabi frequency of the block is
\begin{equation}
  \Omega_n=\sqrt{\Delta_n^2+\lambda^2(n+1)}.
\end{equation}
The short-time contrast of a two-level oscillation started in
$|n,\downarrow\rangle$ is
\begin{equation}
  \mathcal{C}_n=\frac{\lambda^2(n+1)}{\lambda^2(n+1)+\Delta_n^2}.
\end{equation}
At $d=0$, $n=0$, and $\chi$ measured in units of $\lambda$,
the $(\hat a^\dagger\hat a)^2$ form gives
$\Delta_0/\lambda=2f+\chi$ and
\begin{equation}
  \mathcal{C}=\frac{1}{1+(2f+\chi)^2},
\end{equation}
which is Eq.~\eqref{eq:lorentz}. The alternative Kerr
$\hat a^{\dagger 2}\hat a^2$ gives $\Delta_0=2f\lambda$ at every
$\chi$, so it cannot cancel the vacuum residual. That is why only
$(\hat a^\dagger\hat a)^2$ is used for production numbers.

The JC block on $\{|n,\uparrow\rangle,|n+1,\downarrow\rangle\}$ has
the same Kerr shifts with the opposite sign in the detuning,
$\Delta_n^{\rm JC}=\lambda d-\chi D_n$, and no residual at $d=0$.

\section{Excitation-number conservation}
\label{sec:appendix}

For $r=0$, $K=a$. Define $\overline N=a^\dagger a+\sigma_-\sigma_+$.
The free and Kerr terms commute with $\overline N$. For the
interaction,
\begin{align}
[a^\dagger a,\,a^\dagger\sigma_++a\sigma_-]
&=a^\dagger\sigma_+-a\sigma_-,\\
[\sigma_-\sigma_+,\,a^\dagger\sigma_++a\sigma_-]
&=-a^\dagger\sigma_++a\sigma_-,
\end{align}
so $[\overline N,H]=0$ and the space splits into blocks
$\{|n,\downarrow\rangle,|n+1,\uparrow\rangle\}$. This is the $r=0$
solvability of Refs.~\cite{Omolo2021,Omolo2025QIP,Mayero2023,Mayero2024}.

The interaction written in $K$ conserves
$\widetilde{\overline N}=K^\dagger K+\sigma_-\sigma_+$. The
laboratory-frame free term does not:
\begin{equation}
\begin{aligned}
K^\dagger K&=\cosh(2r)\,a^\dagger a
+\sinh r\cosh r\bigl(e^{i\phi}(a^\dagger)^2\\
&\;+e^{-i\phi}a^2\bigr)
+\sinh^2 r,
\end{aligned}
\end{equation}
hence $[a^\dagger a,K^\dagger K]\neq 0$ at $r\neq 0$. The Kerr term
$(\hat n)^2$ likewise fails to commute with $K^\dagger K$. Truncated
full-space diagonalization is therefore required. The $r=0$ theory is
the special case $K=a$.

\section{Truncation}
\label{app:num}

Table~\ref{tab:conv} is the single-atom cutoff walk at $r=1$.
The three production cases change by less than
$1.5\times 10^{-4}$ between $n_{\max}=160$ and $200$.
Table~\ref{tab:convdicke} is the Dicke walk at $\chi=0$.
Cells with $f\le 0.5$, and $N=8$ at $f=5$, settle at
$n_{\max}=160$; Table~\ref{tab:chi0} uses those values.
At $f\ge 1$, except $N=8$ at $f=5$, $t_\ast$ in
Table~\ref{tab:convdicke} jumps when $n_{\max}$ is raised.
Those Table~\ref{tab:chi0} entries are two-digit bounds.
The inequality $\mathcal{C}_{N=16}(f=5)\gg 0.138$ holds in
every column of Table~\ref{tab:convdicke}.

\begin{table}[h]
\centering
\footnotesize
\caption{Short-time contrast versus $n_{\max}$ at $r=1$,
single atom.}
\label{tab:conv}
\begin{tabular}{@{}rcccccc@{}}
\toprule
 & \multicolumn{2}{c}{$f=0.2$, $\chi=-0.4$}
 & \multicolumn{2}{c}{$f=2$, $\chi=0$}
 & \multicolumn{2}{c}{$f=1$, $\chi=-0.3$} \\
\cmidrule(lr){2-3}\cmidrule(lr){4-5}\cmidrule(lr){6-7}
$n_{\max}$ & $\mathcal{C}$ & $\Delta$ & $\mathcal{C}$ & $\Delta$ & $\mathcal{C}$ & $\Delta$ \\
\midrule
$80$  & $0.4765$ & --- & $0.3022$ & --- & $0.3343$ & --- \\
$120$ & $0.4792$ & $2.8\times 10^{-3}$ & $0.2991$ & $3.1\times 10^{-3}$ & $0.3402$ & $5.9\times 10^{-3}$ \\
$160$ & $0.4797$ & $4.2\times 10^{-4}$ & $0.2987$ & $3.6\times 10^{-4}$ & $0.3410$ & $8.0\times 10^{-4}$ \\
$200$ & $0.4797$ & $6.8\times 10^{-5}$ & $0.2989$ & $1.4\times 10^{-4}$ & $0.3411$ & $5.8\times 10^{-5}$ \\
\bottomrule
\end{tabular}
\end{table}

\begin{table}[t]
\centering
\footnotesize
\caption{Short-time Dicke contrast versus $n_{\max}$ at $r=1$,
$\chi=0$.}
\label{tab:convdicke}
\begin{tabular}{@{}rrcccccc@{}}
\toprule
 & & \multicolumn{2}{c}{$n_{\max}=80$}
 & \multicolumn{2}{c}{$120$}
 & \multicolumn{2}{c}{$160$} \\
\cmidrule(lr){3-4}\cmidrule(lr){5-6}\cmidrule(lr){7-8}
$N$ & $f$ & $\mathcal{C}$ & $t_\ast$ & $\mathcal{C}$ & $t_\ast$ & $\mathcal{C}$ & $t_\ast$ \\
\midrule
$8$  & $0.05$ & $0.983$ & $1.925$ & $0.983$ & $1.925$ & $0.983$ & $1.925$ \\
$8$  & $0.20$ & $0.947$ & $1.555$ & $0.947$ & $1.555$ & $0.947$ & $1.555$ \\
$8$  & $0.50$ & $0.887$ & $1.330$ & $0.891$ & $1.370$ & $0.892$ & $1.380$ \\
$8$  & $1.00$ & $0.846$ & $1.195$ & $0.876$ & $2.160$ & $0.896$ & $2.105$ \\
$8$  & $2.00$ & $0.805$ & $0.925$ & $0.898$ & $1.620$ & $0.864$ & $1.415$ \\
$8$  & $5.00$ & $0.572$ & $0.565$ & $0.576$ & $0.555$ & $0.575$ & $0.550$ \\
$16$ & $0.05$ & $0.994$ & $1.640$ & $0.993$ & $1.615$ & $0.993$ & $1.615$ \\
$16$ & $0.20$ & $0.981$ & $1.365$ & $0.979$ & $1.335$ & $0.978$ & $1.330$ \\
$16$ & $0.50$ & $0.954$ & $1.175$ & $0.952$ & $1.175$ & $0.953$ & $1.185$ \\
$16$ & $1.00$ & $0.923$ & $1.065$ & $0.947$ & $1.745$ & $0.934$ & $1.605$ \\
$16$ & $2.00$ & $0.910$ & $0.960$ & $0.878$ & $0.860$ & $0.962$ & $1.450$ \\
$16$ & $5.00$ & $0.820$ & $0.795$ & $0.796$ & $0.725$ & $0.782$ & $0.710$ \\
\bottomrule
\end{tabular}
\end{table}

\end{appendices}

\bibliographystyle{sn-mathphys-num}



\end{document}